# THE PSEUDO-HYPERBOLIC FUNCTIONS AND THE MATRIX REPRESENTATION OF EISENSTEIN COMPLEX NUMBERS

G. DATTOLI

ENEA - Dipartimento Tecnologie Fisiche e Nuovi Materiali
Centro Ricerche Frascati, Roma

E. SABIA

ENEA - Tecnologie Fisiche e Nuovi Materiali
Centro Ricerche Portici, Napoli

M. DEL FRANCO

OSPITE ENEA

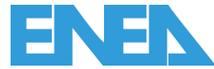

# THE PSEUDO-HYPERBOLIC FUNCTIONS AND THE MATRIX REPRESENTATION OF EISENSTEIN COMPLEX NUMBERS

G. DATTOLI

ENEA - Dipartimento Tecnologie Fisiche e Nuovi Materiali, Centro Ricerche Frascati

E. SABIA

ENEA - Dipartimento Tecnologie Fisiche e Nuovi Materiali, Centro Ricerche Portici, Napoli

M. DEL FRANCO - OSPITE ENEA

# THE PSEUDO-HYPERBOLIC FUNCTIONS AND THE MATRIX REPRESENTATION OF EISENSTEIN COMPLEX NUMBERS

**Abstract**

*We consider the matrix representation of the Eisenstein numbers and in this context we discuss the theory of the Pseudo Hyperbolic Functions. We develop a geometrical interpretation and show the usefulness of the method in Physical problems related to the anomalous scattering of light by crystals*

**Riassunto**

In questo lavoro si utilizza la rappresentazione matriciale dei numeri di Einstein in relazione alla teoria delle funzioni pseudo iperboliche. Tale approccio permette una interpretazione geometrica di tali funzioni di cui si discute l'importanza in merito alla diffusione anomala della luce da parte dei cristalli.

**Parole chiave: Numeri di Einstein e Harwitz, funzioni pseudo iperboliche, diffusione anamala, cristalli**

# INDICE

# THE PSEUDO-HYPERBOLIC FUNCTIONS AND THE MATRIX REPRESENTATION OF EISENSTEIN COMPLEX NUMBERS

## 1. INTRODUCTION

In this paper we will discuss the theory of pseudo-hyperbolic function (PHF) [1] starting from the properties of the matrix representation of Eisenstein and Hurwitz complex numbers [2].

In a previous paper [3] Migliorati, Ricci, and one of the present authors (G. D.), have discussed the link between the Eisenstein numbers and PHF, by showing that such a point of view is the natural frame in which the theory of these functions can be developed. Here we will generalize the point of view of ref. [3], include the Hurwitz complex numbers and show that the matrix representation offers a powerful tool providing a more insightful understanding of their properties and the possibility of extending their use in crystallographic studies.

Just to start with a very familiar example, we consider the $2\times 2$ matrix

$$\hat{h} = \begin{pmatrix} 0 & 1 \\ 1 & 0 \end{pmatrix} \tag{1}$$

usually called the **_hyperbolic_** unit matrix, linked to the unit matrix by the identity

$$\hat{h}^2 = \hat{1} \tag{2}$$

The exponentiation of $\hat{h}$ yields

$$\hat{R}_h(\alpha) = e^{\alpha \hat{h}} = \sum_{n=0}^{\infty} \frac{\alpha^n}{(2n)!} \hat{1} \sum_{n=0}^{\infty} \frac{\alpha^{n+1}}{(2n+1)!} \hat{h} = \begin{pmatrix} \cosh(\alpha) & \sinh(\alpha) \\ \sinh(\alpha) & \cosh(\alpha) \end{pmatrix} \qquad (3).$$

representing a rotation in the hyperbolic space.

All the properties of the hyperbolic functions (HF) can be inferred from those of the matrix (1) and from its exponentiation.

We note, indeed, that

$$\det({}_h\hat{R}(\alpha)) = e^{Tr(\hat{h})} = 1 \qquad (4)$$

which implies the fundamental relation $\cosh^2(\alpha) - \sinh^2(\alpha) = 1$.

The addition theorems are furthermore a consequence of the identity

$$\hat{R}_h(\alpha + \beta) = e^{(\alpha+\beta)\hat{h}} = \hat{R}_h(\alpha) \circ \hat{R}_h(\beta) \qquad (5).$$

The hyperbolic unit is a kind of *"imaginary"* unit, and can be viewed as one of the square roots of the unit matrix, we can therefore consider the following $3\times 3$ matrix generalizations

$$_3\hat{h} = \begin{pmatrix} 0 & 1 & 0 \\ 0 & 0 & 1 \\ 1 & 0 & 0 \end{pmatrix}, \quad _3\hat{k} = \begin{pmatrix} 0 & 0 & 1 \\ 1 & 0 & 0 \\ 0 & 1 & 0 \end{pmatrix} \qquad (6)$$

which satisfy the identities

$$\begin{aligned} _3\hat{h}^2 &= {}_3\hat{k}, \; _3\hat{h}^3 = \hat{1}, \\ _3\hat{h}^+ &= {}_3\hat{k} \end{aligned} \qquad (7)$$

and which are clearly recognized as associated with the cubic roots of the unit matrix.

The exponentiation of $_3\hat{h}$ is slightly more complicated than a simple hyperbolic rotation.

The use of the identities (7) and of the cyclical properties of the successive powers of the matrix $_3\hat{h}$, yields

$$_3\hat{R}_h(\alpha) = e^{\alpha\ _3\hat{h}} = e_0(\alpha)\hat{1} + e_1(\alpha)\ _3\hat{h} + e_2(\alpha)\ _3\hat{k} =$$

$$= \begin{pmatrix} e_0(\alpha) & e_1(\alpha) & e_2(\alpha) \\ e_2(\alpha) & e_0(\alpha) & e_1(\alpha) \\ e_1(\alpha) & e_2(\alpha) & e_0(\alpha) \end{pmatrix}, e_k(\alpha) = \sum_{r=0}^{\infty} \frac{\alpha^{3r+k}}{(3r+k)!}, k = 0,1,2, \tag{8a}$$

and

$$\sum_{k=0}^{2} e_k(\alpha) = e^{\alpha} \tag{8b}$$

The functions $e_k(\alpha)$ are a generalization of the hyperbolic functions and are the already quoted PHF [1], whose properties can be obtained in full analogy to the case of HF. The condition on the determinant of $_3\hat{R}_h(\alpha)$ provides the fundamental identity of PHF

$$\det(_3\hat{R}_h(\alpha)) = e^{Tr(_3\hat{h})} = 1 \Rightarrow$$
$$\Rightarrow e_0^3(\alpha) + e_1^3(\alpha) + e_2^3(\alpha) - 3e_0(\alpha)e_1(\alpha)e_2(\alpha) = 1 \tag{9}$$

furthermore, the use of the elementary identity

$$a^3 + b^3 + c^3 - 3abc = (a+b+c)\left(a^2 + b^2 + c^2 - ab - bc - ac\right) \tag{10}$$

and of eqs. (8,9), provides us with the further relation

$$e_0^2(\alpha) + e_1^2(\alpha) + e_2^2(\alpha) - e_0(\alpha)e_1(\alpha) - e_1(\alpha)e_2(\alpha) - e_0(\alpha)e_2(\alpha) = e^{-\alpha} \tag{11}$$

As for the ordinary HF, the addition theorems follow from the properties of the matrix product

$$_3\hat{R}_h(\alpha+\beta) = e^{(\alpha+\beta)\ _3\hat{h}} = {_3\hat{R}_h(\alpha)} \circ {_3\hat{R}_h(\beta)} \Rightarrow$$
$$\Rightarrow e_k(\alpha+\beta) = e_k(\alpha)e_k(\beta) + e_{k+1}(\alpha)e_{k+2}(\beta) + e_{k+2}(\alpha)e_{k+1}(\beta). \tag{12}$$
$$e_{k+3n}(\alpha) = e_k(\alpha), k = 0,1,2$$

Furthermore, by keeping repeated derivatives of both sides of eq. (8a), with respect to $\alpha$, we find that the PHF satisfy the differential equation [1]

$$\partial_\alpha^k e_s(\alpha) = e_{s+2k}(\alpha),$$
$$s = 0,1,2 \quad k = 1,2,3$$
(13).

In the following sections we will develop a geometrical interpretation of the above functions.

## 2. THE TRI-COMPLEX NUMBERS AND THE PHF

The ordinary hyperbolic functions can be geometrically interpreted as reported in Fig. 1, where they are defined with reference to the unit hyperbola (which plays the same role of the unit circle, in the case of the ordinary trigonometric functions) and the sector area, defining the argument of the HF.

As already remarked the matrix $\hat{R}_h(\alpha)$ represents a non Euclidean rotation which preserves the norm $x^2 - y^2 = 1$.

In the following we will show that an analogous geometric interpretation holds for the PHF too.

To this aim we introduce a *"tri-complex number"* in terms of $3 \times 3$ matrices as follows [4]

$$\hat{\zeta} = x\hat{1} + {}_3\hat{h}\, y + {}_3\hat{k}\, z = \begin{pmatrix} x & y & z \\ z & x & y \\ y & z & x \end{pmatrix}$$
(14).

which is expressed in terms of the matrices $\{\hat{1}, {}_3\hat{h}, {}_3\hat{k}\}$ which are all commuting and therefore *x, y, z* are coplanar. The determinant of the above matrix is

$$\det(\hat{\zeta}) = x^3 + y^3 + z^3 - 3xyz \tag{15}.$$

and it is not difficult to understand that the *"number"* $\hat{\zeta}$ can be expressed in terms of the PHF, as it follows

$$\hat{\zeta} = e^{\beta} e^{\gamma \, _3\hat{h}}, \tag{16}$$

accordingly we find

$$\begin{aligned} x &= e^{\beta} e_0(\gamma), \\ y &= e^{\beta} e_1(\gamma), \\ z &= e^{\beta} e_2(\gamma) \end{aligned} \tag{17}.$$

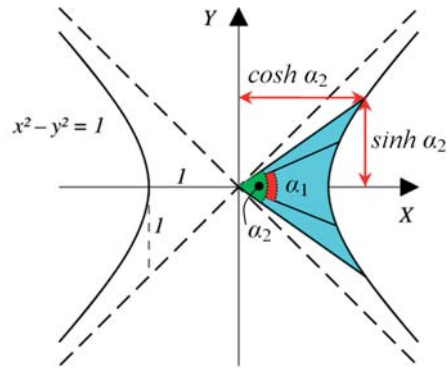

*Fig. 1 - Sector area and hyperbolic functions*

The coefficients $\beta, \gamma$ can be expressed in terms of $x, y, z$ by noting that (see eqs. (8))

$$\begin{aligned} \det(\hat{\zeta}) &= e^{3\beta} \det\left|_3\hat{R}(\gamma)\right| = e^{3\beta} = x^3 + y^3 + z^3 - 3xyz, \\ x + y + z &= e^{\beta}\left(e_0(\gamma) + e_1(\gamma) + e_2(\gamma)\right) = e^{\beta + \gamma} \end{aligned} \tag{18a},$$

therefore we find

$$\begin{aligned} e^{\beta} &= \sqrt[3]{x^3 + y^3 + z^3 - 3xyz}, \\ e^{\gamma} &= \frac{x + y + z}{\sqrt[3]{x^3 + y^3 + z^3 - 3xyz}} \end{aligned} \tag{18b}.$$

We can therefore conclude that the action of the matrix $_3\hat{R}_h(\alpha)$ on the tri-complex number can be written as

$$_3\hat{R}_h(\alpha)\hat{\zeta} = e^{\beta} \, _3\hat{R}_h(\alpha + \gamma) \tag{19},$$

which is amenable for an interesting geometrical interpretation, which will be discussed in the forthcoming section.

## 3. THE GEOMETRY OF PHF

We have already stressed that $_3\hat{h}$ is a root of the unit matrix and it can therefore be viewed as a representation of the Eisenstein unit [2], defined as

$$\hat{\omega} = -\frac{1 - i\sqrt{3}}{2} = e^{i\frac{2\pi}{3}} \tag{20a}$$

and specified by the properties

$$\hat{\omega}^2 + \hat{\omega} = -1,$$
$$\hat{\omega}^3 = 1 \tag{20b}$$

The complex quantity

$$\rho = a + \hat{\omega} b \tag{21}$$

defines a Eisenstein number, whose geometric representation is given in Fig. 2

The Eisenstein complex form a Euclidean domain, whose norm is defined through the identity

$$(a + \hat{\omega} b)(a + \hat{\omega}^2 b) = a^2 - ab + b^2 \tag{22}$$

The norm given in eq. (22) specifies the modulus of the vector reported in Fig. 3, which is the sum of the vectors with moduli $a,b$, forming an angle $2\pi/3$.

The transformation preserving such a norm is easily obtained by noting that it is associated with the modulus of the complex number

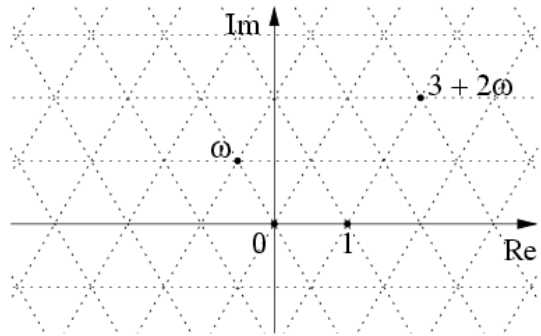

*Fig. 2 - The lattice of Eisenstein numbers*

$$\rho = a + \hat{\omega} b = (a - \frac{1}{2}b) + i\frac{\sqrt{3}}{2}b \qquad (23)$$

which in matrix form writes

$$\rho = \begin{pmatrix} 1 & -\frac{1}{2} \\ 0 & \frac{\sqrt{3}}{2} \end{pmatrix} \begin{pmatrix} a \\ b \end{pmatrix} \qquad (24)$$

The norm of the vector $\rho$ is then conserved by the **pseudo-rotation**

$$\hat{R}(\alpha, \frac{\pi}{6}) = \begin{pmatrix} \cos(\alpha) & -\sin(\alpha) \\ \sin(\alpha) & \cos(\alpha) \end{pmatrix} \begin{pmatrix} 1 & -\frac{1}{2} \\ 0 & \frac{\sqrt{3}}{2} \end{pmatrix} = \begin{pmatrix} \cos(\alpha) & -\sin(\alpha + \frac{\pi}{6}) \\ \sin(\alpha) & \cos(\alpha + \frac{\pi}{6}) \end{pmatrix} \qquad (25).$$

The tri-complex number introduced in eq. (14) can be realized in terms of the Eisenstein unit as

$$\zeta = x + y\hat{\omega} + z\hat{\omega}^2 = \left[x - \frac{1}{2}(y+z)\right] - i\frac{\sqrt{3}}{2}(y-z) \qquad (26).$$

and in polar notation can be written as

$$\zeta = |\zeta|e^{i\varphi},$$
$$|\zeta| = \sqrt{x^2 + y^2 + z^2 - xy - xz - zy} \qquad (27).$$
$$\tan(\varphi) = \frac{\sqrt{3}(y-z)}{2x - (y+z)}$$

The modulus of $\zeta$ can be geometrically interpreted as shown in Fig. 3, furthermore, by noting that

$$x^3 + y^3 + z^3 - 3xyz = (x + y + z)|\zeta|^2 \qquad (28)$$

which, on account of eq. (18b), yields

$$e^\beta = \sqrt[3]{|\zeta|^2|v|},$$

$$e^\gamma = \sqrt[3]{\frac{|v|^2}{|\zeta|^2}}, \qquad (29).$$

$$|v| = x + y + z$$

We can understand the geometrical meaning of $|v|$ by proceeding as follows.

We start from eq. (26) and use the standard procedure of orthonormalization to introduce the following ortho-normal triple [4]

$$\underline{\xi} = \frac{1}{\sqrt{6}} \begin{pmatrix} 2 & -1 & -1 \\ 0 & \sqrt{3} & -\sqrt{3} \\ \sqrt{2} & \sqrt{2} & \sqrt{2} \end{pmatrix} \begin{pmatrix} x \\ y \\ z \end{pmatrix} \qquad (30)$$

reported in Fig. 3. The angle $\varphi$ lies in the plane individuated by the axes $\xi_1, \xi_2$ and we can interpret $|v|$ as linked to the length $\overline{OO'} = |v|/\sqrt{3}$.

The matrix ${}_3\hat{R}_h(\alpha)$ in the space of the Eisenstein numbers can be written as

$${}_3\hat{R}_h(\alpha) \equiv e^{\hat{\omega}\alpha} = e_0(\alpha) + \hat{\omega} e_1(\alpha) + \hat{\omega}^2 e_2(\alpha) = e^{-\frac{\alpha}{2}} e^{i\frac{\sqrt{3}}{2}\alpha} \qquad (31).$$

we can therefore write

$${}_3\hat{R}_h(\alpha)\hat{\zeta} \equiv |\zeta| e^{-\frac{\alpha}{2}} e^{i(\varphi - \frac{\sqrt{3}}{2})\alpha} \qquad (32)$$

and it can be geometrically interpreted as planar rotation of the vector $\xi$ from $\varphi$ to $\varphi - \sqrt{3}/2\ \alpha$ with a reduction of its modulus by a factor $e^{-\alpha/2}$.

In conclusion we get that the matrix

$$\hat{E}(\alpha) = e^{\frac{\alpha}{2}} {}_3\hat{R}_h(\alpha) \qquad (33)$$

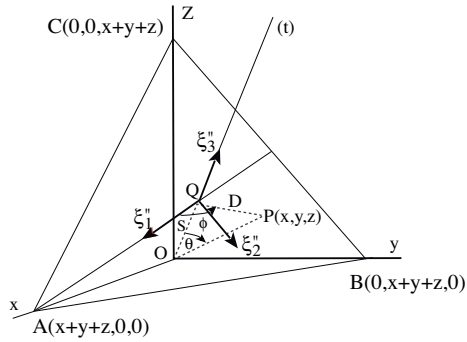

*Fig. 3 - Geometry of tri-complex numbers*

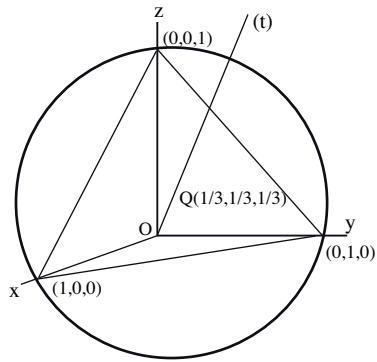

*Fig. 4 - Invariant circle of tri-complex number*

induces transformation which preserves the modulus of the tri-complex number $\hat{\zeta}$, which for $|\zeta| = 1$, is defined on the invariant circle reported in Fig. 4 (see also ref. (4)).

## 4. CONCLUDING REMARKS

The properties of the matrix associated with the Eisenstein numbers are general enough to allow further generalizations of the properties of tricomplex numbers and PHF we have discussed so far.

Before going further we recall that the two variable Hermite polynomials [5]

$$H_n(x,y) = n! \sum_{r=0}^{[n/2]} \frac{x^{n-2r} y^r}{(n-2r)! r!} \tag{34}$$

can be defined by means of the generating function

$$\sum_{n=0}^{\infty} \frac{t^n}{n!} H_n(x,y) = e^{xt+yt^2} \tag{35}.$$

The following exponential

$$_3\hat{R}_h(\alpha,\eta) = e^{\alpha\ _3\hat{h}+\eta\ _3\hat{k}} = e^{\alpha\ _3\hat{h}+\eta\ _3\hat{h}^2} \tag{36}$$

containing either *h and k* matrices, can be written as

$$_3\hat{R}_h(\alpha,\eta) = \sum_{n=0}^{\infty} \frac{_3\hat{h}^n}{n!} H_n(\alpha,\eta) = {}_he_0(\alpha,\eta)\hat{1} + {}_he_1(\alpha,\eta)\ _3\hat{h} + {}_he_2(\alpha,\eta)\ _3\hat{k},$$

$$_he_s(\alpha,\eta) = \sum_{r=0}^{\infty} \frac{H_{3r+s}(\alpha,\eta)}{(3r+s)!}, s = 0,1,2 \tag{37}$$

the two variable functions $_he_s(\alpha,\beta)$ represent a generalization of the PHF, which, on account of the following property of the two variable Hermite polynomials [5]

$$e^{y\partial_x^2} x^n = H_n(x,y) \tag{38}$$

can also be written as

$$_he_s(\alpha,\eta) = e^{\eta\partial_\alpha^2} e_s(\alpha) \tag{39}.$$

The action of the matrix $_3\hat{R}_h(\alpha,\eta)$ on the tri-complex number $\hat{\zeta}$ can therefore be written as

$$_3\hat{R}_h(\alpha,\eta)\hat{\zeta} \equiv |\zeta| e^{-\frac{\alpha}{2}} e^{i(\varphi - \frac{\sqrt{3}}{2}\alpha) + \eta(\frac{1+i\sqrt{3}}{2})^2} =$$

$$= |\zeta| e^{-(\frac{\alpha+\eta}{2})} e^{i\left[\varphi + \frac{\sqrt{3}}{2}(\eta-\alpha)\right]} = e^{-\frac{\eta}{2}}\ _3\hat{R}(\alpha-\eta)\hat{\zeta} \tag{40}.$$

It is well known that the symmetry generated by the roots of unity is widely exploited in crystallography [6]. Here we will consider the transformation which can be generated by the point operators [7], given in Tab. 1.

**Tab. 1 – Matrix representation of the crystallographic group**

$$\hat{R}_1 = \begin{pmatrix} 1 & 0 & 0 \\ 0 & 1 & 0 \\ 0 & 0 & 1 \end{pmatrix}, \ \hat{R}_2 = \begin{pmatrix} -1 & 0 & 0 \\ 0 & -1 & 0 \\ 0 & 0 & 1 \end{pmatrix}, \ \hat{R}_3 = \begin{pmatrix} 1 & 0 & 0 \\ 0 & -1 & 0 \\ 0 & 0 & -1 \end{pmatrix}, \ \hat{R}_4 = \begin{pmatrix} -1 & 0 & 0 \\ 0 & 1 & 0 \\ 0 & 0 & -1 \end{pmatrix},$$

$$\hat{R}_5 = \begin{pmatrix} 0 & 0 & 1 \\ 1 & 0 & 0 \\ 0 & 1 & 0 \end{pmatrix}, \ \hat{R}_6 = \begin{pmatrix} 0 & 0 & -1 \\ -1 & 0 & 0 \\ 0 & 1 & 0 \end{pmatrix}, \ \hat{R}_7 = \begin{pmatrix} 0 & 0 & 1 \\ -1 & 0 & 0 \\ 0 & -1 & 0 \end{pmatrix}, \ \hat{R}_8 = \begin{pmatrix} 0 & 0 & -1 \\ 1 & 0 & 0 \\ 0 & -1 & 0 \end{pmatrix},$$

$$\hat{R}_9 = \begin{pmatrix} 0 & 1 & 0 \\ 0 & 0 & 1 \\ 1 & 0 & 0 \end{pmatrix}, \ \hat{R}_{10} = \begin{pmatrix} 0 & 1 & 0 \\ 0 & 0 & -1 \\ -1 & 0 & 0 \end{pmatrix}, \ \hat{R}_{11} = \begin{pmatrix} 0 & -1 & 0 \\ 0 & 0 & 1 \\ -1 & 0 & 0 \end{pmatrix}, \ \hat{R}_{12} = \begin{pmatrix} 0 & -1 & 0 \\ 0 & 0 & -1 \\ 1 & 0 & 0 \end{pmatrix}$$

The matrices $\hat{R}_{1,\ldots,4}$ induces just reflections, and it is evident that $\hat{R}_5 = {}_3\hat{h}, \hat{R}_9 = {}_3\hat{k}$, it can also be easily verified that the matrices $\hat{R}_6, \hat{R}_{11}$ and $\hat{R}_7, \hat{R}_{12}$ can be considered representations of the cubic roots of unity, since

$$\hat{R}_6^2 = \hat{R}_{11}, \quad \hat{R}_6^3 = \hat{R}_1 = \hat{1}, \quad \hat{R}_7^2 = \hat{R}_{12}, \quad \hat{R}_7^3 = 1 , \tag{41}$$

The extension of the above results to higher order matrices can be easily accomplished, we note indeed that the matrices

$$_4\hat{h} = \begin{pmatrix} 0 & 1 & 0 & 0 \\ 0 & 0 & 1 & 0 \\ 0 & 0 & 0 & 1 \\ 1 & 0 & 0 & 0 \end{pmatrix}, \ _4\hat{k} = {}_4\hat{h}^2 = \begin{pmatrix} 0 & 0 & 1 & 0 \\ 0 & 0 & 0 & 1 \\ 1 & 0 & 0 & 0 \\ 0 & 1 & 0 & 0 \end{pmatrix}, \ _4\hat{l} = {}_4\hat{h}^3 = \begin{pmatrix} 0 & 0 & 0 & 1 \\ 1 & 0 & 0 & 0 \\ 0 & 1 & 0 & 0 \\ 0 & 0 & 1 & 0 \end{pmatrix} \tag{42}$$

represent the fourth roots of unity and can be exploited to define the PHF family

$$e^{\alpha \, _4\hat{h}} = {}_4e_0(\alpha)\hat{1} + {}_4e_1(\alpha) \, _4\hat{h} + {}_4e_2(\alpha) \, _4\hat{k} + {}_4e_3(\alpha) \, _4\hat{l},$$

$$_4e_s(\alpha) = \sum_{r=0}^{\infty} \frac{\alpha^{4r+s}}{(4r+s)!}, s = 0,1,2,3 \tag{43}$$

The same procedure as before can be therefore used to gain a geometrical interpretation of this family of functions.

Before closing this note, it is worth stressing that the use of the generalized Hermite polynomials, can be useful to study exponentializations of the type

$$\hat{E}(\alpha,\eta,\delta) = e^{\alpha \, _4\hat{h} + \eta \, _4\hat{k} + \delta \, _4\hat{l}} \tag{44},$$

which on account of the generating function [5]

$$\sum_{n=0}^{\infty} \frac{t^n}{n!} H_n^{(3)}(x,y,z) = e^{xt + yt^2 + zt^3},$$

$$H_n^{(3)}(x,y,z) = n! \sum_{r=0}^{[n/3]} \frac{z^{n-3r} H_r(x,y)}{(n-3r)! r!} \tag{45}$$

can be written as

$$\hat{E}(\alpha,\eta,\delta) = e_0(\alpha,\eta,\delta)\hat{1} + e_1(\alpha,\eta,\delta) \, _4\hat{h} + _4e_2(\alpha,\eta,\delta) \, _4\hat{k} + e_3(\alpha,\eta,\delta) \, _4\hat{l},$$

$$_h e_s(\alpha,\eta,\delta) = \sum_{r=0}^{\infty} \frac{H_{4r+s}^{(3)}(\alpha,\eta,\delta)}{(4r+s)!}, s = 0,1,2,3 \tag{46}.$$

The polynomials $H_n^{(3)}(x,y,z)$ are three variable Hermite polynomials and the relationship between this family of polynomials and the hypercomplex numbers will be more carefully discussed in a forthcoming investigation.